\documentclass[runningheads]{llncs}
\usepackage[T1]{fontenc}

\usepackage{graphicx}
\usepackage[final]{microtype}
\usepackage{booktabs}
\usepackage[colorlinks=true, allcolors=blue]{hyperref}

\usepackage[sort, compress]{cite}
\usepackage[table]{xcolor}
\usepackage{colortbl}

\begin{document}
%
\title{Beyond MR Image Harmonization:\\ Resolution Matters Too}


%
\author{Savannah~P.~Hays\inst{1}\orcidID{0009-0005-8711-1356}\and
Samuel~W.~Remedios\inst{2}\orcidID{0000-0001-8634-8128} \and
Lianrui~Zuo\inst{3}\orcidID{0000-0002-5923-9097} \and
Ellen~M.~Mowry\inst{4}\orcidID{0000-0003-0623-5188} \and
Scott~D.~Newsome\inst{4}\orcidID{0000-0002-5284-4681} \and
Peter~A.~Calabresi\inst{4}\orcidID{0000-0002-7776-6472} \and
Aaron~Carass\inst{1}\orcidID{0000-0003-4939-5085} \and
Blake~E.~Dewey\inst{4}\orcidID{0000-0003-4554-5058} \and
Jerry~L.~Prince\inst{1}\orcidID{0000-0002-6553-0876}}

\authorrunning{S.P. Hays et al.}
\institute{
Image Analysis and Communications Laboratory,\\
Department of Electrical and Computer Engineering,\\ Johns Hopkins University, Baltimore, MD 21218, USA\\[0.4em]
\and
Department of Computer Science, Johns Hopkins University, Baltimore,~MD~21218,~USA\\[0.4em] \and
Department of Electrical and Computer Engineering,\\ Vanderbilt University, Nashville,~TN~37235,~USA\\[0.4em] \and
Department of Neurology, Johns Hopkins School of Medicine,\\ Baltimore, MD 21287, USA 
}

%
\maketitle              
\begin{abstract}
Magnetic resonance~(MR) imaging is commonly used in the clinical setting to non-invasively monitor the body.
There exists a large variability in MR imaging due to differences in scanner hardware, software, and protocol design.
Ideally, a processing algorithm should perform robustly to this variability, but that is not always the case in reality.
This introduces a need for \textit{image harmonization} to overcome issues of domain shift when performing downstream analysis such as segmentation.
Most image harmonization models focus on acquisition parameters such as inversion time or repetition time, but they ignore an important aspect in MR imaging---resolution.
In this paper, we evaluate the impact of image \textit{resolution} on harmonization using a pretrained harmonization algorithm.
We simulate 2D acquisitions of various slice thicknesses and gaps from 3D acquired, 1mm$^3$ isotropic MR images and demonstrate how the performance of a state-of-the-art image harmonization algorithm varies as resolution changes.
We discuss the most ideal scenarios for image resolution including acquisition orientation when 3D imaging is not available, which is common for many clinical scanners.
Our results show that harmonization on low-resolution images does not account for acquisition resolution and orientation variations.
Super-resolution can be used to alleviate resolution variations but it is not always used.
Our methodology can generalize to help evaluate the impact of image acquisition resolution for multiple tasks.
Determining the limits of a pretrained algorithm is important when considering preprocessing steps and trust in the results.

\end{abstract}
\section{Introduction}
Magnetic resonance~(MR) imaging is a commonly used non-invasive medical imaging modality.
The flexibility of MR imaging lies in the selection of pulse sequences and scanning parameters allowing for the acquisition of multiple MR tissue contrasts that reveal different tissue properties~\cite{prince2006medical}.
Due to this large variation in images that can be acquired and lack of standardization and consistency between images acquired in studies, the domain shift problem~\cite{biberacher2016intra,he2020self,zuo2021unsupervised,zuo2021information} is commonly observed even when precautions are taken during acquisition~\cite{clark2023neuroimaging}.
There have been many efforts to standardize acquisition such as the MAGNIMS-CMSC-NAIMS 2021 MRI guidelines~\cite{wattjes20212021}, but compliance with these guidelines is variable.

Domain shift poses a substantial challenge for downstream analysis such as segmentation, necessitating image harmonization~\cite{zuo2022disentangle, zuo2022haca3, zuo2024deepbook} techniques to mitigate these issues~\cite{carass2024nir}.
Image harmonization typically adjusts image contrasts to match a desired domain, enhancing consistency across different MR images.
Despite the advancements in harmonization techniques, the impact of acquisition resolution on harmonization performance has not been extensively studied.

It is usually preferred that MR images be acquired with 3D pulse sequences, but these protocols generally take longer, often have poorer signal-to-noise-ratios, and do not offer a full range of desired contrasts.
As a result, clinical MR images are often acquired as stacked 2D slices, leading to anisotropic volumes with varying in-plane and through-plane resolutions.
This variability in contrast and resolution can significantly affect the quality and interpretability of MR images.
Contrast refers to the difference in signal intensity between different tissues, which allows for the differentiation of various tissue types.
Resolution, on the other hand, refers to the ability to distinguish small structures within the image.
Although they are distinct concepts, contrast and resolution are interrelated.
For instance, high in-plane resolution with poor through-plane resolution results in images with different contrast characteristics between planes due to the partial volume effect, complicating both human interpretation and automated analysis.
Slice thickness leads to partial volume effects, which is present in all images, and is made worse with increasing slice thickness.
Slice gap at a naive level is the distance between two adjacent slices.
This is a gap of missing data that cannot be recovered in general.
In addition to partial volume effects, another unwanted artifact that can cause differences in the appearance of anatomy is aliasing, which is the overlapping of frequency components resulting from a sample rate below the Nyquist rate.
These overlaps lead to distortions when the signal is reconstructed.
It can be very difficult to undo these distortions.

\begin{figure}[!tb]
   \centerline{\includegraphics[width = 0.8\columnwidth]{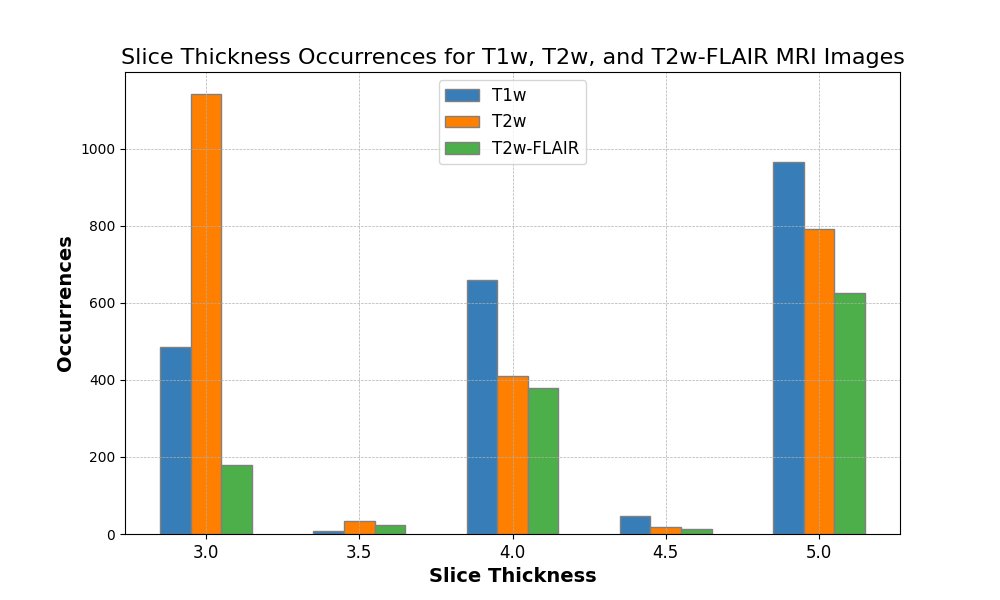}}
   \caption{Slice thickness occurrences for each MRI image contrast: T1w, T2w, and T2w-FLAIR.}
   \label{fig:thickness_stats} 
   
\end{figure}

The importance of addressing these issues is underscored by a large, nationwide pragmatic study of people with MS~(PwMS) (NCT0350032), where significant variability in acquisition parameters was observed despite protocol guidelines~\cite{dewey2021improving,hays2023quantifying}.
The pragmatic trial involves approximately 50 imaging sites nationwide and over 800 PwMS, and despite guidelines that requested 3D acquisitions, there was a prevalence of 2D acquisitions in the scans, resulting in high variability in image resolution and contrast. 
Out of 17,056 imaging volumes, only 5,652 were 3D acquired, while the remaining 11,404 volumes were 2D acquired with varying slice thicknesses and orientations as shown in Fig.~\ref{fig:thickness_stats}, leading to inconsistent image quality.
Out of the 5,652 3D acquired volumes, 3,413 are T1-weighted~(T1w) images, 151 are T2-weighted~(T2w) images, and 2,088 are T2w Fluid-Attenuated Inversion Recovery~(T2w-FLAIR) images.
Out of the 11,404 2D acquired volumes, 5,832 are T1w images, 3,458 are T2w images, and 2,114 are T2w-FLAIR images.
The acquired orientation for each image contrast was heavily favored to be axial for all 3 contrasts.
There were approximately 1,020 sagittal scans and only 184 coronal scans out of all of the image contrasts.

Trained radiologists can interpret and diagnose from these anisotropic volumes despite the variations in contrast and resolution.
However, automated volumetric processing techniques often require isotropic voxels for optimal performance.
Many super-resolution~(SR) algorithms have been proposed to map low-resolution~(LR) anisotropic volumes to desired high-resolution~(HR) isotropic volumes~\cite{remedios2023sashimi, remedios2024spie, wu2024mlmi}.
These algorithms enhance image quality by improving both resolution and contrast, making the images more suitable for both human interpretation and algorithmic analysis.
Despite their effectiveness, SR algorithms are not always incorporated into preprocessing pipelines for tasks such as harmonization.

Currently, Harmonization with Attention-based Contrast, Anatomy, and Artifact Awareness~(HACA3)~\cite{zuo2022haca3} is employed for image harmonization in a pragmatic study of PwMS.
HACA3 is an open source\footnote{\url{https://github.com/lianruizuo/haca3}} unsupervised image harmonization approach for structural MR images.
It does not require paired subject data for training, distinguishing it from methods like DeepHarmony~\cite{dewey2019deepharmony}.
Its encoder-decoder structure learns latent representations of anatomy, acquisition contrast, and image quality, enabling harmonization by combining these representations effectively.
Despite its advanced capabilities, HACA3’s performance in relation to different acquisition resolutions and orientations needs thorough evaluation given its use on diverse clinical data.
HACA3 has the flexibility to handle multiple source MR images when harmonizing.
The attention mechanism helps distinguish which source MR images the network should focus on based on the source MR image quality and similarity to the target MR image contrast.
It was shown to have superior performance when multiple source MR images are used given they are included in the training contrasts.

In this work, we evaluate HACA3’s performance based on image acquisition resolution and orientation.
Our study aims to identify the most ideal scenarios for using HACA3 when 3D imaging is unavailable, which is common in many clinical protocols.
Our results indicate that simple interpolation methods like cubic B-spline do not consistently improve harmonization performance as slice thickness increases, reinforcing the necessity of SR to alleviate the burden of 2D acquired images effectively.
By addressing the limitations of current harmonization approaches and emphasizing the need for resolution-aware preprocessing steps, our work provides crucial insights into improving MR image analysis, ensuring more reliable and accurate results in clinical settings.


\section{Methods}
\subsection{Data and Preprocessing}
Our dataset involves 10 subjects, each with three image contrasts: T1w, T2w, and T2w-FLAIR.
All images are originally 3D acquired with 1mm$^3$ isotropic resolution and bias field corrected using N4~\cite{tustison2010n4itk}.
To simulate 2D acquisitions, we degraded these HR images to various LR images.
We simulated LR by first blurring the HR images with a Shinnar-Le Roux slice selection profile (which is commonly done in true MR acquisition~\cite{pauly1991parameter, ikonomidou2000improved}) followed by downsampling by the slice separation with cubic B-spline interpolation. We used the implementation of the Shinnar-Le Roux algorithm provided by \texttt{sigpy}~\cite{martin2020sigpy}.
Since HACA3 expects images registered to a 1mm$^3$ isotropic MNI space, we registered the LR images back to this space after degradation using ANTs~\cite{avants2009ants} and 3D cubic B-spline interpolation.

We degraded the images to eight different resolutions in three orientations (axial, sagittal, and coronal), resulting in 24 different degradations per image contrast.
The image resolution is denoted as ``slice thickness $\|$ slice gap'', where ``$\|$'' can be read as ``skip'' and both numbers are reported in millimeters.
Slice gap is the difference between the slice spacing and the slice thickness.
For example, $3\|0$ means a slice thickness of 3mm with no gap.
The evaluated resolutions include: $3\|0$ (highest resolution), $3\|1$, $4\|0$, $4\|1$, $4\|1.2$, $5\|0$, $5\|1$, and $5\|1.5$ (lowest resolution).

\subsection{Harmonization with HACA3}
For testing HACA3, we used three image contrasts (T1w, T2w, and T2w-FLAIR) as input.
While HACA3 has the capability for contrast dropout, it performs best with all available contrasts.
We evaluated different combinations of LR images as inputs to HACA3 to determine how performance varies with image resolution.
Our harmonization target was representative of images acquired at a local scanner following the imaging protocol for a pragmatic study of PwMS, as our in-house image processing pipeline uses this contrast harmonization target for lesion segmentation.

We conducted several experiments to assess the performance of HACA3 with different acquisition resolutions and orientations.
With multiple tissue contrasts the hope is that HACA3 can convert these signals into a better anatomical signal.
The experiments are summarized in Table~\ref{tab:experiments}.
Experiments~1a--1c investigate when all input images to HACA3 are 2D acquired with the same orientation (axial, sagittal, or coronal).
Experiment~2 investigates when all input images to HACA3 are 2D acquired, but with different orientations (axial T1w, sagittal T2w, and coronal T2w-FLAIR).
Experiments~3 and~4 investigate when the T1w image is 3D acquired, but the T2w and T2w-FLAIR images are 2D acquired with the same (axial) and different orientations, respectively.
Experiments~5 and~6 investigate when the T2w-FLAIR image is 3D acquired, but the T1w and T2w images are 2D acquired with the same (axial) or different orientations, respectively.
We measured the peak signal-to-noise ratio~(PSNR) and structural similarity index measure~(SSIM) for the harmonized images against the ground truth, which are the harmonized images obtained using all 3D, 1mm$^3$ isotropic images as input to HACA3.


\begin{table}[!tb]
    \rowcolors{2}{cyan!10}{white}
    \centering
    \caption{Simulated acquisition resolution and orientation for different experiments.}
    \label{tab:experiments}
    \renewcommand{\arraystretch}{1.25} 
    \setlength{\tabcolsep}{1ex} 
    \begin{tabular}{ c c rl c rl c rl}
    \toprule
    \textbf{Experiment} &\hspace*{3ex}& \multicolumn{2}{c}{\textbf{T1w}} && \multicolumn{2}{c}{\textbf{T2w}} && \multicolumn{2}{c}{\textbf{T2w-FLAIR}} \\
    \cmidrule(lr){1-10}
    %
    1a && 2D & Axial &\hspace*{3ex}& 2D & Axial &\hspace*{3ex}& 2D & Axial \\
    1b && 2D & Sagittal && 2D & Sagittal && 2D & Sagittal \\
    1c && 2D & Coronal && 2D & Coronal && 2D & Coronal \\
    2 && 2D & Axial && 2D & Sagittal && 2D & Coronal \\
    3 && \multicolumn{2}{c}{3D} && 2D & Axial && 2D & Axial \\
    4 && \multicolumn{2}{c}{3D} && 2D & Sagittal && 2D & Coronal \\
    5 && 2D & Axial && 2D & Axial && \multicolumn{2}{c}{3D} \\
    6 && 2D & Sagittal && 2D & Coronal && \multicolumn{2}{c}{3D} \\
    \bottomrule
    \end{tabular}
\end{table}

\section{Experiments and Results}

\begin{figure}[!tb]
   \centerline{\includegraphics[width = 0.9\columnwidth]{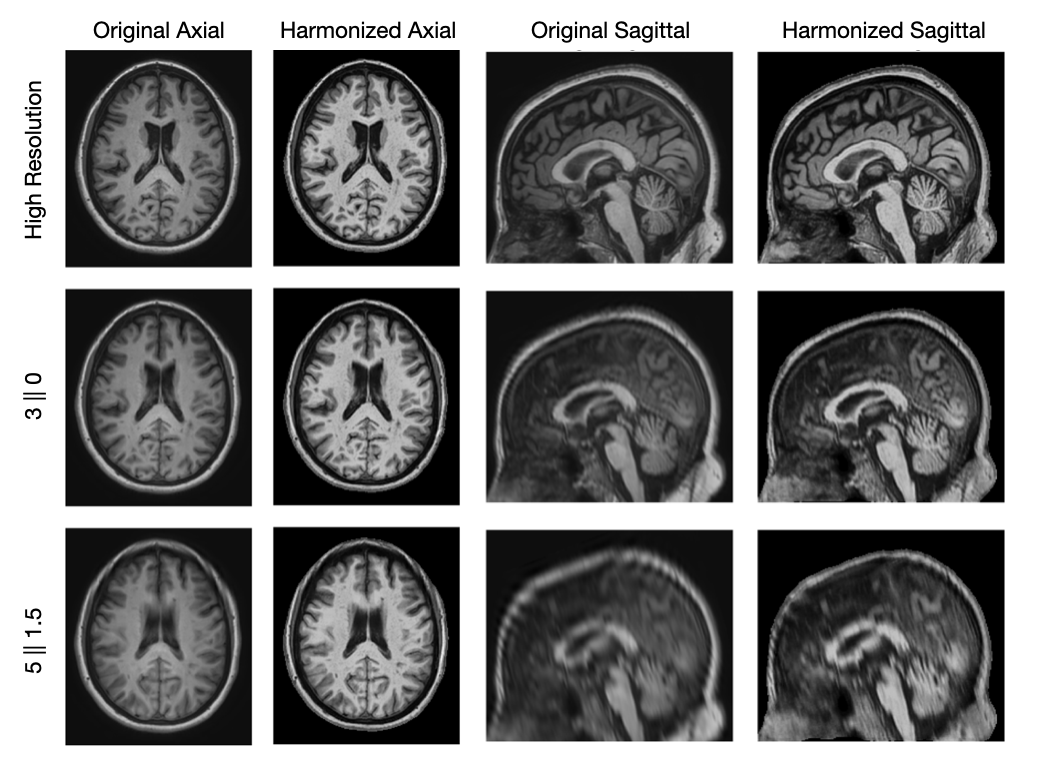}}
   \caption{The original and harmonized registered T1w images at HR and two LRs, all images are shown with the same look-up table.}
   \label{fig:t1_images} 
\end{figure}

Figure~\ref{fig:t1_images} shows the original and harmonized registered T1w images at HR and two simulated LRs demonstrating the impact of resolution degradation.
We show a LR image simulating an axial acquisition at a resolution of $3\|0$, which is the best 2D resolution we investigate and a LR image simulating a sagittal acquisition at a resolution of $5\|1.5$, which is the worst resolution we investigate.
As the resolution gets worse, we observe that the image appears more blurry due to the slice thickness.
The T2w and T2w-FLAIR images behave similarly to the T1w images.
   
\begin{figure}[!tbh]
   \centerline{\includegraphics[width = 0.9\columnwidth]{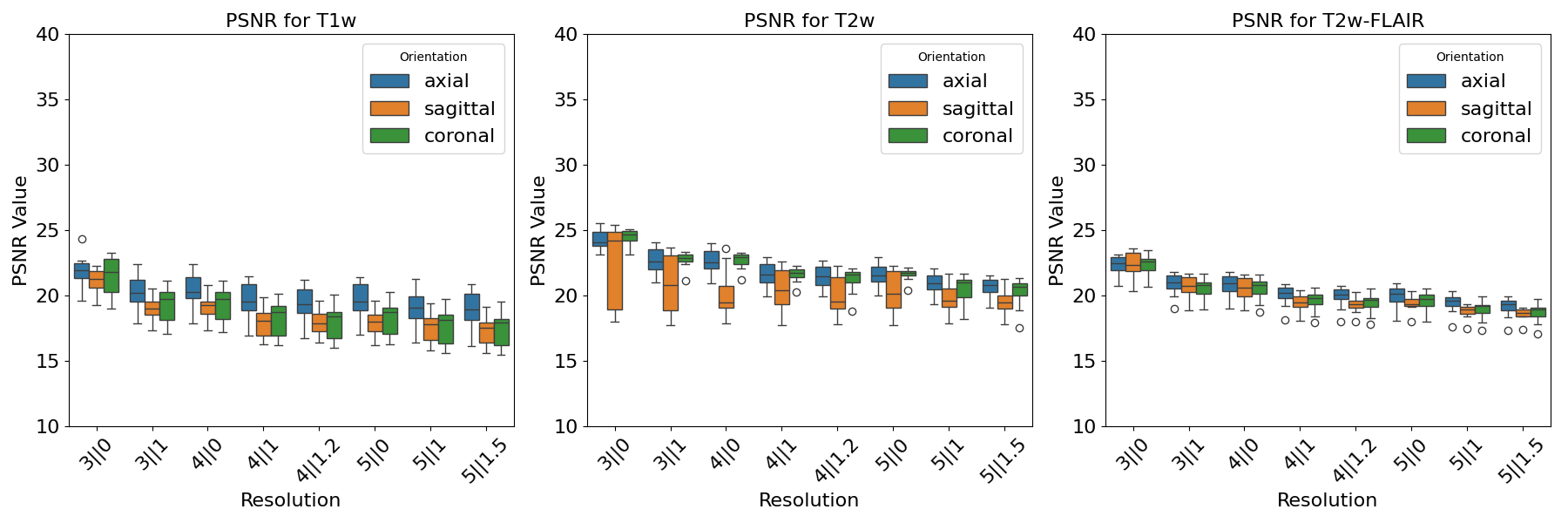}}
   \caption{\textbf{Experiment~1a--c:} PSNR when all input images are 2D acquired with the same orientation.}
   \label{fig:all_2D_same} 
\end{figure}

\begin{figure}[!tbh]
   \centerline{\includegraphics[width = 0.9\columnwidth]{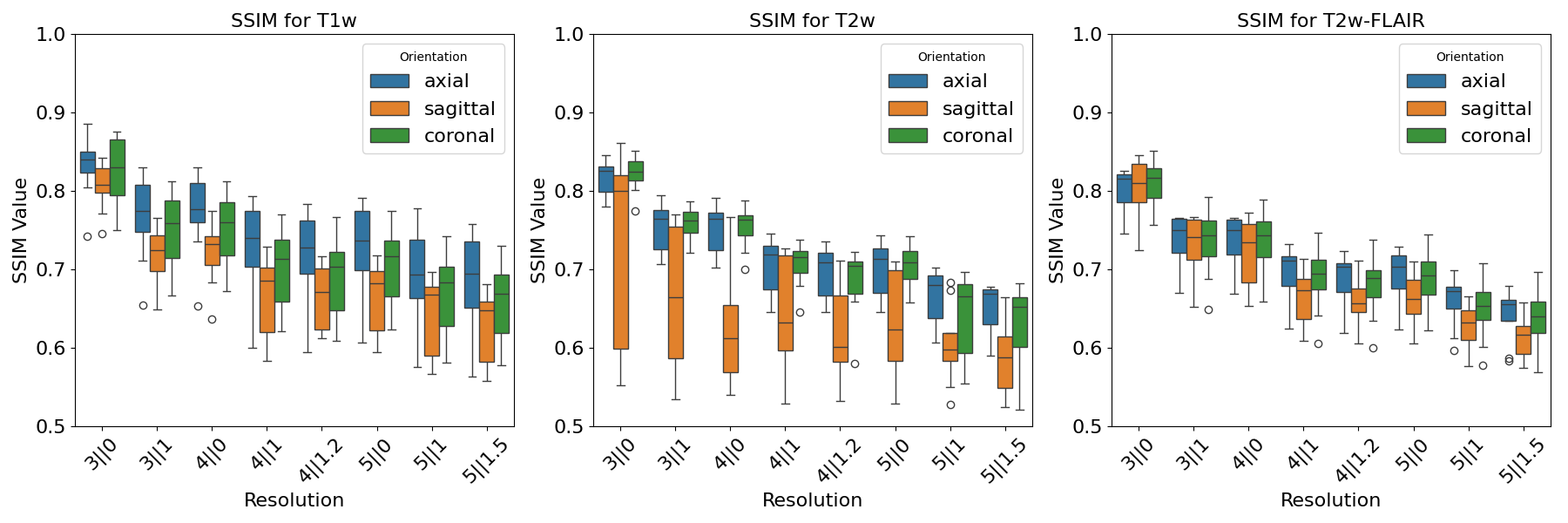}}
   \caption{\textbf{Experiment~1a--c:} SSIM when all input images are 2D acquired with the same orientation.}
   \label{fig:ssim_all_2D_same} 
\end{figure}

In Experiment~1a-c, all input images are 2D acquired with the same orientation (axial, sagittal, or coronal).
The PSNR and SSIM results for these experiments are shown in Figs.~\ref{fig:all_2D_same}~and~\ref{fig:ssim_all_2D_same}.
We synthesized the harmonized images using HACA3 with the input images simulated with the same orientation and resolution.
These experiments show that HACA3 performs consistently when all input images are acquired with the same orientation and that the harmonized image quality is best when the input images are axial acquisitions.

\begin{figure}[!tbh]
   \centerline{\includegraphics[width = 0.9\columnwidth]{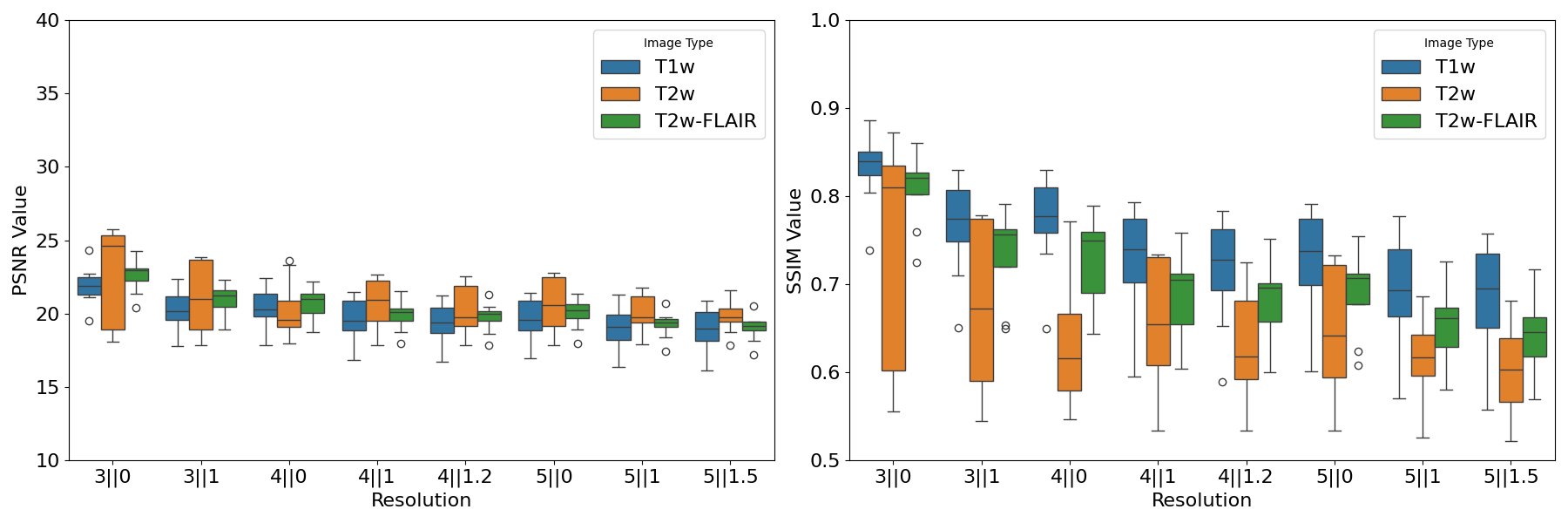}}
   \caption{\textbf{Experiment~2:} PSNR and SSIM when all images are 2D acquired with different orientations (T1w: axial, T2w: sagittal, and T2w-FLAIR: coronal).}
   \label{fig:all_2D_diff} 
\end{figure}

In Experiment~2, the input images are 2D acquired with different orientations (T1w: axial, T2w: sagittal, and T2w-FLAIR: coronal).
    The PSNR and SSIM for this experiment are shown in Fig.~\ref{fig:all_2D_diff}.
This experiment show that having 2D images acquired with different orientations does not improve HACA3 performance and that the variation in orientations negatively impacts the harmonization quality compared to the same-orientation acquisitions.

\begin{figure}[!tbh]
   \centerline{\includegraphics[width = 0.9\columnwidth]{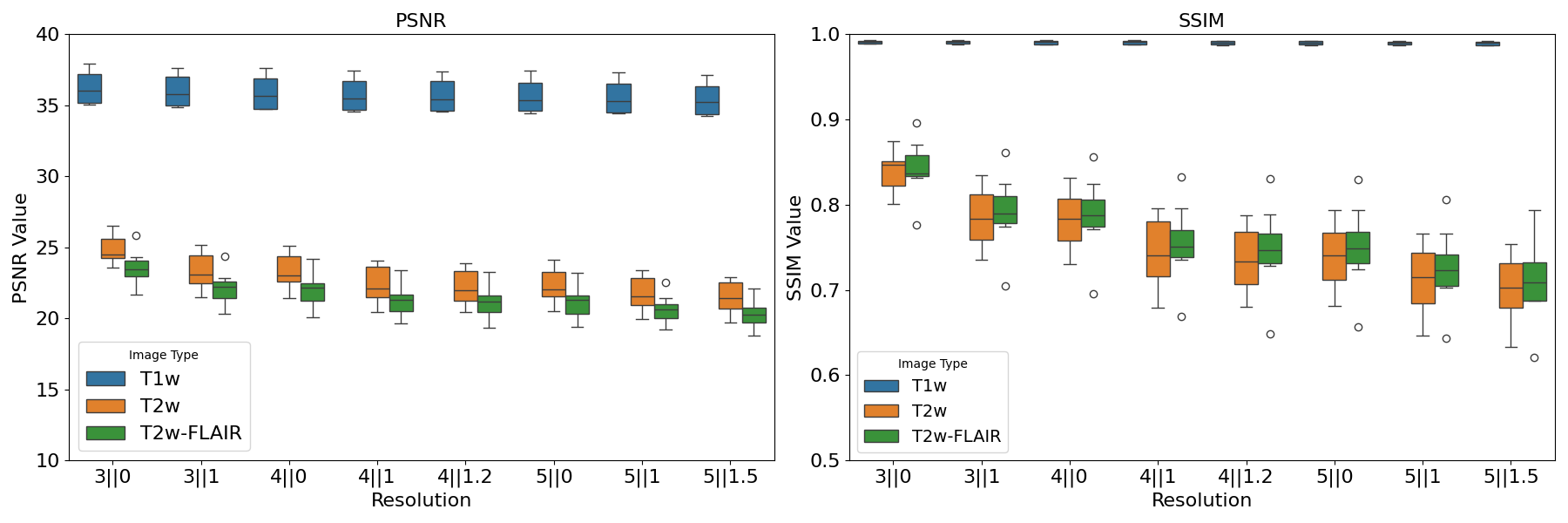}}
   \caption{\textbf{Experiment~3:} PSNR and SSIM when the T1w image is 3D acquired and the T2w and T2w-FLAIR images are 2D acquired with the same orientation (axial).}
   \label{fig:T1w3d_T2waxial_flaxial} 
\end{figure}

\begin{figure}[!tbh]
   \centerline{\includegraphics[width = 0.9\columnwidth]{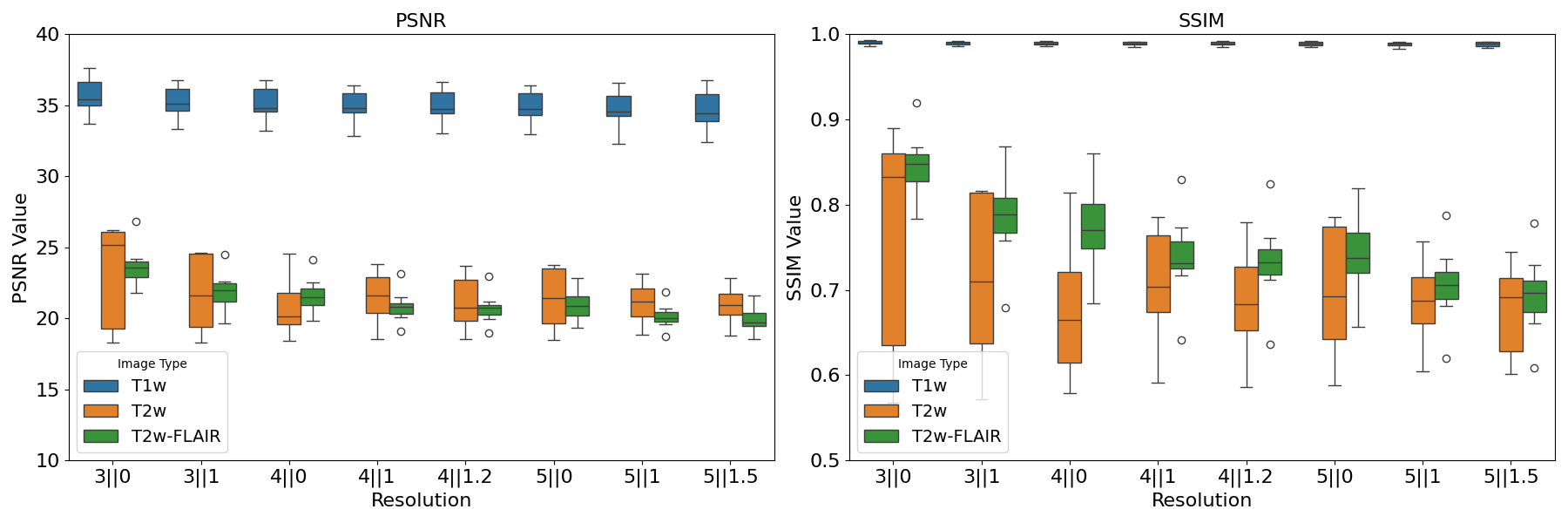}}
   \caption{\textbf{Experiment~4:} PSNR and SSIM when the T1w image is 3D acquired and the T2w and T2w-FLAIR images are 2D acquired with different orientations.}
   \label{fig:T1w3d_T2wsag_flcor} 
\end{figure}

In Experiments~3 and~4, the input T1w image is 3D acquired, while the T2w and T2w-FLAIR images are 2D acquired with the same orientation (axial) and different orientations, respectively.
The PSNR and SSIM for these experiments are shown in Fig.~\ref{fig:T1w3d_T2waxial_flaxial}~and~\ref{fig:T1w3d_T2wsag_flcor}.
These experiments show that using a 3D T1w image with 2D T2w and T2w-FLAIR images with the same orientation provides better harmonization quality while the mixed orientations do not improve performance.

\begin{figure}[!tbh]
   \centerline{\includegraphics[width = 0.9\columnwidth]{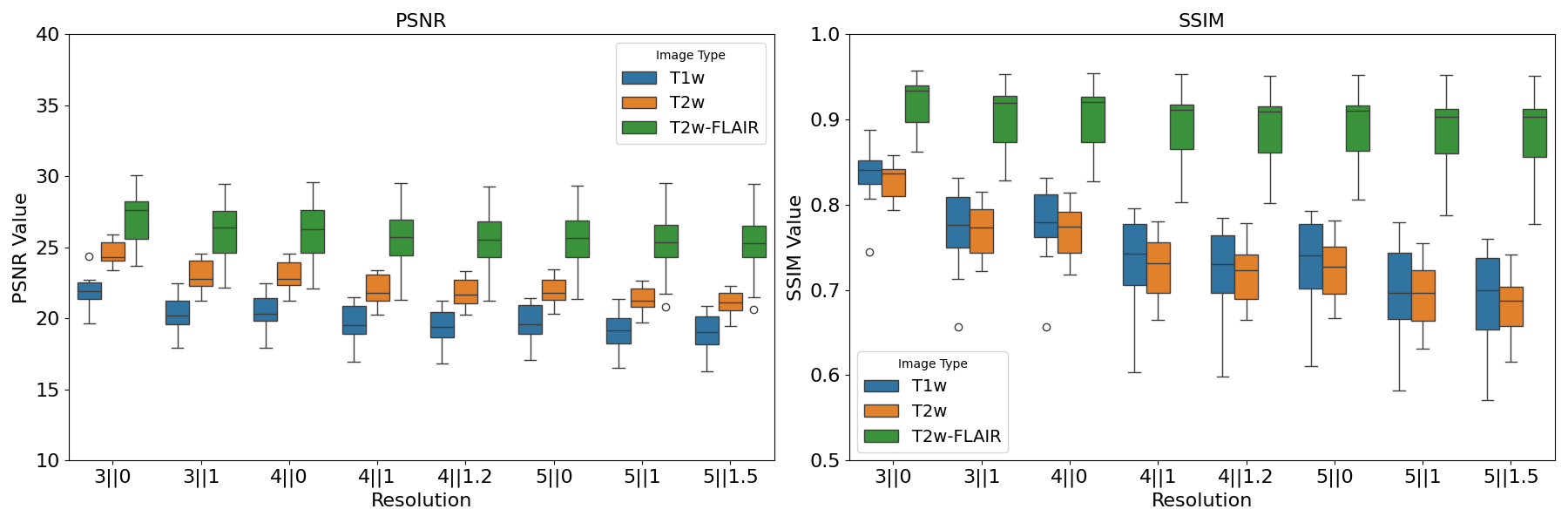}}
   \caption{\textbf{Experiment~5:} PSNR and SSIM for HACA3 performance when the input T2w-FLAIR image is 3D acquired and T1w and T2w images are 2D acquired at the same orientation (axial).}
   \label{fig:T1waxial_T2waxial_fl3d} 
\end{figure}

In Experiments~5 and~6, the input T2w-FLAIR image is 3D acquired and the T1w and T2w images are 2D acquired with the same orientation (axial) and different orientations, respectively.
The PSNR and SSIM for Experiment~5 is shown in Fig.~\ref{fig:T1waxial_T2waxial_fl3d}.
These experiments show that using a 3D T2w-FLAIR image with 2D T1w and T2w images with the same orientation results in better harmonization performance while the mixed orientations do not have a major impact on HACA3 performance.

From these experiments, HACA3 performs best with consistent, high-resolution, same-orientation acquisitions.
Mixed orientations do not noticeably improve harmonization quality.
The use of 3D acquisitions for at least one image contrast enhances performance for that particular image contrast but does not aid in synthesizing the other contrasts.

\section{Discussion and Conclusion}
Although 3D acquisition is ideal, it remains absent from many clinical protocols.
The variability in clinical MR images often leads to inconsistent processing reliability and accuracy.
Image harmonization is commonly used to address domain shift issues in deep learning-based algorithms; however, its performance can vary with image resolution.

In this study, we evaluated the performance of the state-of-the-art MR image harmonization algorithm, HACA3, across different acquired resolutions.
Our findings indicate the impact of orientation and resolution, the effect of 3D and 2D image combinations, and HACA3 limitations.
When sagittal acquisitions are used, there is a notable variance in HACA3 performance, particularly for T2w image synthesis (Fig.~\ref{fig:all_2D_same}).
Performance declines sharply as resolution decreases.
Similar performance patterns are observed for different slice spacings, such as $3\|1$ and $4\|0$ as well as $4\|1$ and $5\|0$.
Surprisingly, including one 3D T1w image does not enhance performance for the other 2D image contrasts~(Fig.~\ref{fig:T1w3d_T2waxial_flaxial}).
The attention mechanism in HACA3 is effective at recognizing differences in contrast variations but does not account for differences in resolution.
This limitation affects the algorithm's performance when dealing with input images of varying resolutions.
In experiments with sagittal T2w and coronal T2w-FLAIR images~(Fig.~\ref{fig:T1w3d_T2wsag_flcor}), HACA3's performance did not improve for the T2w-FLAIR synthesis, but there was increased variance for T2w synthesis at the higher resolutions.
This suggests that HACA3 does not perform optimally with LR sagittal acquisitions.
HACA3 was unable to super-resolve the 2D images that were registered to a 1mm$^3$ isotropic MNI space, as expected.
HACA3 aims to preserve anatomy while only changing contrast, maintaining the observed anatomy's resolution.
Some HACA3 performance degradation could be attributed to registration errors, which are common when processing LR acquisitions.
While 2D acquired images were part of HACA3's training data, they were all super-resolved during preprocessing.
However, SR is not always performed on 2D acquired images, and its effectiveness decreases with increasing slice thickness.
SR is also computationally expensive.

In this work, we presented a comprehensive evaluation of the impact of MR image resolution on the HACA3 harmonization algorithm.
We highlighted the variability of 2D acquisitions in a longitudinal, clinical study of PwMS.
Our results demonstrate that as slice spacing increases, HACA3 performance degrades.
This finding underscores that harmonization alone does not solve the problem when using LR images.
While we cannot control the imaging scanners available to patients, it is crucial to manage the resolution used for scanning when HR images are not acquired.
Implementing SR techniques may be necessary to enhance the performance of harmonization algorithms like HACA3, ensuring more reliable and accurate results in clinical settings.

A promising direction for future work is to focus on the attention mechanism in HACA3.
Training HACA3 to recognize and adapt to resolution differences, in addition to contrast variations, could improve its performance when input images with varying resolutions are used.
This advancement would make the harmonization process more robust and adaptable to the diverse conditions encountered in clinical practice.


\subsection*{Acknowledgments}
\label{sect:acknolwedgments}
This material is partially supported by the Johns Hopkins University Percy Pierre Fellowship~(Hays) and the National Science Foundation Graduate Research Fellowship under Grant No. DGE-2139757~(Hays) and Grant No. DGE-1746891~(Remedios). Development is partially supported by FG-2008-36966~(Dewey), CDMRP W81XWH2010912~(Prince), NIH R01 CA253923~(Landman), NIH R01 CA275015~(Landman), the National MS Society grant RG-1507-05243~(Pham) and Patient-Centered Outcomes Research Institute~(PCORI) grant MS-1610-37115~(Newsome and Mowry).
The statements in this publication are solely the responsibility of the authors and do not necessarily represent the views of the Patient-Centered Outcomes Research Institute~(PCORI), its Board of Governors or Methodology Committee.

\bibliographystyle{splncs04}
\bibliography{cas-refs}

\end{document}